\documentclass{aa}
\usepackage{graphicx}
\usepackage{txfonts}
\usepackage{natbib}
\bibpunct{(}{)}{;}{a}{}{,}

\begin{document}

  \title{Intermittent nulls in PSR~B0818$-$13, and the subpulse-drift alias mode.}

  \author{Gemma Janssen \and Joeri van Leeuwen}
      
  \institute{Astronomical Institute, Utrecht University, PO Box 80\,000,
     3508 TA Utrecht, The Netherlands}
  
  \offprints{\\Gemma Janssen, \email{g.h.janssen@astro.uu.nl} \\
	Joeri van Leeuwen, \email{a.g.j.vanleeuwen@astro.uu.nl} }
 
  \abstract{We show that all long nulls in PSR~B0818$-$13 are trains
  of rapidly alternating nulls and pulses (each shorter than one pulse
  period). Sometimes only the nulls coincide with our pulse window,
  resulting in one of the apparently long nulls seen occasionally. We
  show these are seen as often as expected if during such a
  train the probability for nulls is 1.2 times less than for pulses.

  During nulls, the subpulse drift-speed appears to increase. We assume
  that the carousel of sparks that possibly underlies the subpulses
  actually slows down, as it does in similar pulsars like
  PSR~B0809+74, and conclude that the subpulse-drift in this pulsar
  must be aliased. The carousel must then rotate in 30 seconds or less,
  making it the fastest found to date.
  }

  \date{Received/Accepted}

\maketitle

\section{Introduction}

\begin{figure*}[t]
  \begin{minipage}[b]{120mm}
  \includegraphics[width=\textwidth]{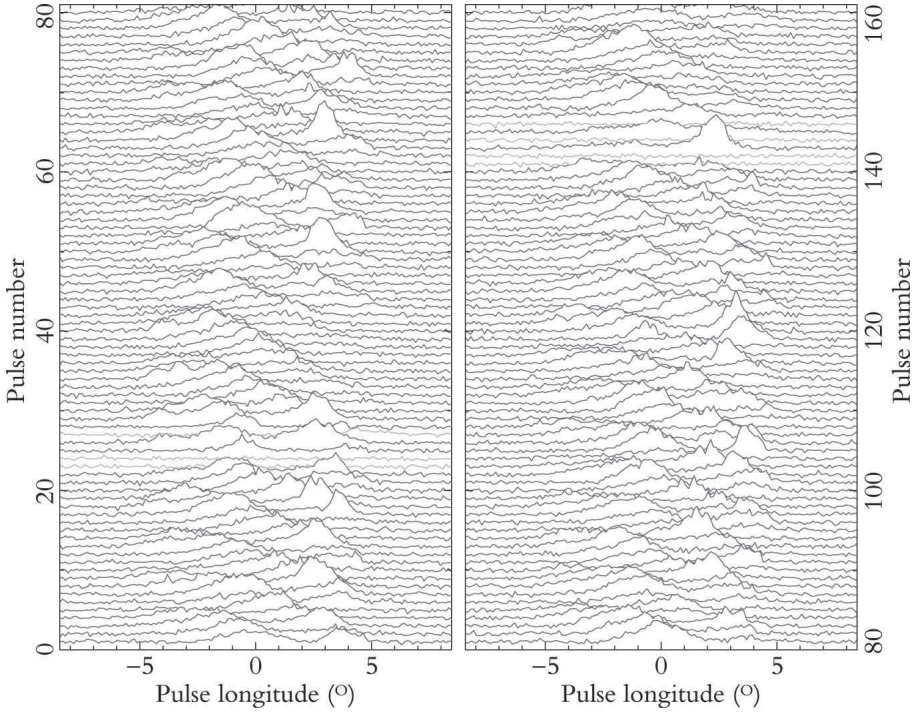}
  \end{minipage}%
  \hspace{5mm}%
  \begin{minipage}[b]{55mm}
  \caption{Stacked single pulses with two intermittent nulls in lighter
  grey.}
  \vspace{1cm}
  \label{img:stack0818}
  \end{minipage}%
\end{figure*}

	Observations of drifting subpulses are a nice example of
undersampling: in some radio pulsars the positions of the pulse
components change regularly (as in fig.\ \ref{img:stack0818}), but as
these positions are determined only once per pulse period, the exact
underlying motion remains unknown.

	Finding this underlying motion, and its relation to other
pulsar parameters, could be helpful in determining which, if any, of
the proposed mechanisms to generate drifting subpulses
\citep{rs75,wri03} is correct. We will assume the subpulses are formed
by discrete locations of emission (`sparks') that rotate around the
magnetic pole of the pulsar, causing the pulse components to move
through the pulse windows (`drifting subpulses'). Different directions
and speeds of this rotation can produce identical subpulse positions
at the observer, so-called aliasing \citep[see also fig.\ 2
in][]{lsr+03}. To determine the actual speed of the spark carousel,
one can check for the periodicity caused as brighter or offset sparks
re-appear after one carousel rotation. \citet{dr99} did so for
PSR~B0943+10, and found a carousel rotation time of around 41 seconds,
longer than expected from theory \citep{rs75}. With the rotation time
known, \citet{dr99} could track individual sparks. They noted that
sparks retain their characteristic brightness and positions for about
100 seconds. If this number were the same in other pulsars, carousel
rotation times longer than 100 seconds could not be found by a
periodicity search, as the characteristic traits of the subpulses will
have changed before they return into view. For such long carousel
rotation periods a different technique is needed.

	In some pulsars there is a mechanism that can be used to
circumvent the undersampling problem and hence determine the
underlying carousel-speed: nulling. In these pulsars, a few percent of
the pulses are missing as the pulsars suddenly turns off
(`nulls'). After these nulls, the subpulse drift is often
affected. Van Leeuwen et~al. (2002, 2003) \nocite{lkr+02, lsr+03}
investigated this drifting-nulling interaction in PSR~B0809+74 and
find it can only be explained if the underlying carousel rotates in
over 200 seconds, again much longer than expected. They predict that
in pulsars with shorter carousel rotation times the drift direction
should reverse after nulls.

In their 1983 paper, Lyne \& Ashworth \nocite{la83} mention the
nulling-drifting interaction of PSR~B0818$-$13 and show the post-null
drift behaviour (their fig.\ 14). For long nulls, there appears to be
a reversed drift direction. Combining this with the prediction of
\citet{lsr+03} we decided to further investigate the
drifting-nulling interaction in PSR~B0818$-$13.

\section{Observations and data reduction}
We have observed PSR~B0818$-$13 from 2000 to 2003 with PuMa, the
Pulsar Machine \citep{vkh+02} at the Westerbork Synthesis Radio
Telescope (WSRT). The 11 observations amount to 20 hours in total or,
using the 1.238\,s period, 5.8$\times$10$^4$ individual pulses. All
observations were conducted at a time resolution of 0.4096\,ms and
around frequencies in between 328 and 382\,MHz, with a bandwidth of
several times 10\,MHz, depending on interference. Each 10\,MHz band was
split in 128 channels and dedispersed offline.

The data were searched for nulls and subpulses in the way described in
\citet{lkr+02}: for each observation we compose a histogram of the
energy of individual pulses, like fig.\ \ref{img:energyhist}. The
discrete population around zero-energy pulses is labeled `nulls'. In
each individual pulse, we fit Gaussian profiles to the subpulses. We
determine which subpulse-fits are significant and store the subpulse
positions, widths and heights along with the raw pulse data. As the
subpulse-drift bands form regular structures, we use the predictive
capability of the thus-found strong subpulses to detect weaker
subpulses over the noise.

\section{Results: analysis of nulls and subpulse drift}
\subsection*{Statistics of short and long nulls}

\begin{figure}[t]
   \includegraphics[]{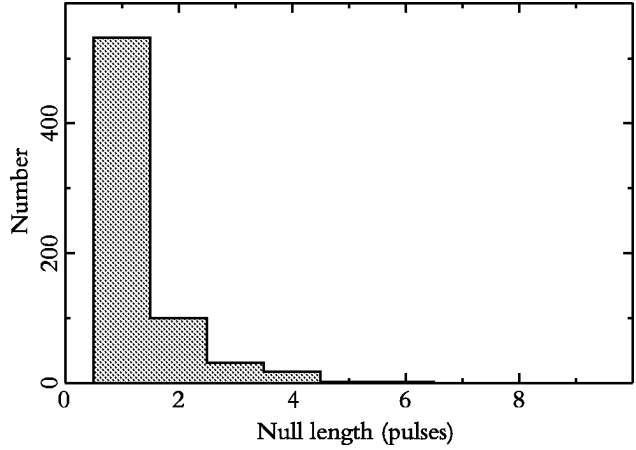}
   \caption{Histogram of the null lengths.}
   \label{img:nullhist}
\end{figure}

PSR~B0818$-$13 nulls about 1\% of the time. As the histogram of null
lengths fig.\ \ref{img:nullhist} shows, most of these nulls are very
short, with lengths of only one or two periods. What the histogram
doesn't show however, is that although roughly 2/3 of these short
nulls occur by themselves, many are observed in groups of intermittent
pulses and nulls, like the groups seen in fig.\ \ref{img:stack0818}.
As this behaviour is uncommon, \citet{la83} chose to disregard all
intermittent short nulls, and only investigate longer nulls. But has
the pulsar really turned off during such a long null? Or are the long
nulls in PSR~B0818$-$13 actually intermittent nulls where the series of
short nulls happen to line up with our line of sight? This is the
question we investigate.

\begin{figure}[t]
   \includegraphics[]{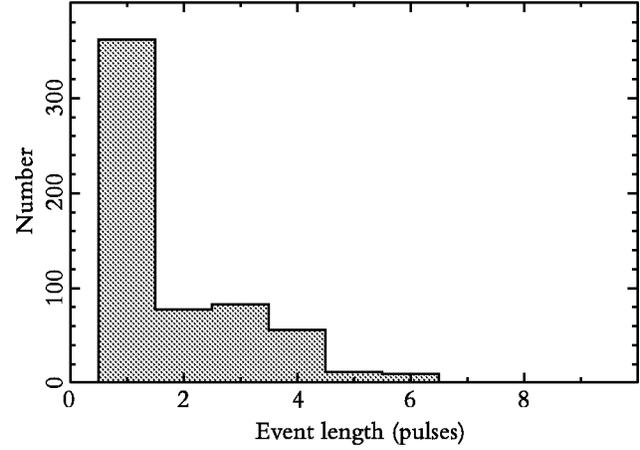}
   \caption{Histogram of the event lengths.}
   \label{img:eventhist}
\end{figure}

If we assume that, during a train, the pulsar quickly and successively
turns on and off, we can use the statistics of the null/pulse
occurrence within a train to determine the likeliness ratio of null
and pulse occurrence. To classify the different pulse trains, we have
defined the first and last nulls of a train to be the start and end of
the group. In fig.\ \ref{img:eventhist} we give an overview of all
event lengths. From this figure it is clear that the occurrence of
null-groups as a whole is not governed by chance: there are too many
long events. Within an event the pulsar blinks fairly quickly, but
the change back to the normal pulse state usually takes several
seconds.

The null group in the left panel of fig.\ \ref{img:stack0818}, for
example, we classify as an event of length 5, while the one in the
right panel is an event of length 6. As the first and last nulls act
as markers, there are only 6$-$2=4 positions in the rightmost of fig.\
\ref{img:stack0818} group 
\begin{figure}[h]
  \includegraphics[]{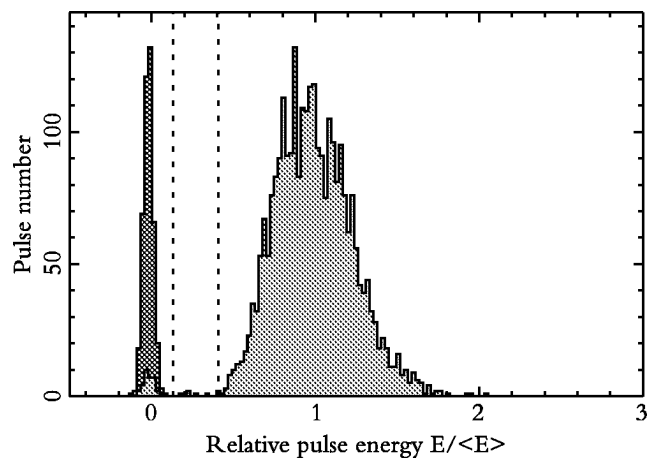}
  \caption{Energy histogram for PSR~B0818$-$13. In light grey we show
  the energy distribution of the energies 
  found in the on-pulse window. In black we show the energies
  found in an equally large off-pulse window, caused by system noise.
  The set of low-energy pulses detached from both the normal and the
  null distribution are probably half-nulls.}
  \label{img:energyhist}
\end{figure}
that can be either nulls or pulses. In this
case, there are two nulls and two pulses in this middle part, in a
NNPNPN pattern.
In table \ref{tab:groups} we give an overview of all
the different null-pulse-trains present in our sample, in the same
pattern notation. In the middle parts of all events, there are 238
possible locations to be divided between pulses and nulls. Of these,
107 are nulls, and 131 are pulses, so we conclude that, on average,
pulses are 1.2 times more likely to occur within a group than
nulls. Using this ratio, we can predict the chance that a train of a
certain length appears to be one long null. If we consider that we
probably miss or misidentify events that start or end with pulses in
stead of nulls, the occurrence of long apparent nulls agrees quite well
with the prediction.  For groups of length 5, for example, the first
and last positions are nulls by definition, but the middle 3 positions
should be nulls, the probability of which is
$(\frac{1.0}{1.0+1.2})^3$=0.09. This is in good agreement with the
observed ratio of 1 in 9 that follows from table \ref{tab:groups}.

This means that at the pulsar the intermittent pulse trains and the
long nulls are the same. When the short nulls in such a train all
happen to occur when the pulsar faces us, the effect only appears
different.

We now know that during the null-trains, the pulsar blinks quickly,
producing several nulls that are shorter than a pulse period. The
continuous increase towards shorter nulls in null-length histogram
fig.\ \ref{img:nullhist} already hinted at the existence of a large
number of nulls shorter than 1 pulse period. Some of the starts and
ends of these nulls must occur within the pulse window. Because of the
variability normally already present in the pulse shapes, this dying
or growing of the flux is not directly recognizable, but statistically
it results in a lower energy for the pulse. In the histogram of pulse
energies these `half-nulls', or `half-pulses' should be located in
between the energy-values of the real nulls and the real pulses (the
vertical lines in fig.\ \ref{img:energyhist}).

In the histogram, we do indeed see a number of these low-energy pulses
(about 0.3\%, with energies varying from 10 to 40\% of the
average pulse energy). For a pulse to have 10-40\% of the normal
energy, the null should start (stop) in the last (first) 20\% of the
pulse window, so within $\pm$ 2$^\circ$. If the positions of the
starts and stops of null are random, we would expect to see this
happen in $\frac{2}{360}$=0.5\% of pulses, in good agreement with the
values found above. If these low-energy pulses are really
half-nulls, we would expect them to occur preferentially near
null trains, as normal nulls do. We therefore calculated the mean
distance to a null for both normal and low-energy pulses. We find that
low-energy pulses are on average about 30\% closer to a null than
normal pulses, so from this we conclude that at least some of the
low-energy pulses are actually half-nulls, where we see the pulsar
turning on or off while facing us.

\begin{table}[b]%
  \centering
  \begin{tabular}{|c|c|c|cc|c|}%
    \hline
    Event  &  pulse--null  & \# of  &       &      & total \\
    length &  combinations & events & \# N  & \# P &       \\
    \hline
    \hline
    1 & N & 361 & & & \\
    2 & NN & 76 & & & \\
    3 & NNN & 27 & 27 & 0 & 27 \\
      & NPN & 52 & 0 & 52 & 52 \\
    4 & NNNN & 16 & 32 & 0 & 32 \\
      & NPPN & 13 & 0 & 26 & 26 \\
      & NPNN & 19 & 19 & 19 & 38 \\
    5 & NNNNN & 1 & 3 & 0 & 3 \\
      & NPPPN & 2 & 0 & 6 & 6 \\
      & NPPNN & 1 & 1 & 2 & 3 \\
      & NPNPN & 1 & 1 & 2 & 3 \\
      & NPNNN & 4 & 8 & 4 & 12 \\
    6 & NNNNNN & 1 & 4 & 0 & 4 \\
      & NPPPNN & 2 & 2 & 6 & 8 \\
      & NPPNPN & 3 & 3 & 9 & 12 \\
      & NPNPNN & 2 & 4 & 4 & 8 \\
      & NPNNNN & 1 & 3 & 1 & 4 \\
    \hline
    \hline
    & total: & & 107 & 131 & 238 \\
    \hline
    & fraction & & .45 & .55 & 1\\
    \hline
  \end{tabular}
  \caption{Length, pattern and occurrence of all null-groups in our
  sample. By definition, events always start and end with nulls. N
  denotes a null, P a pulse. Symmetric patterns are treated as one:
  NPNN for example is a combination of NPNN and NNPN. Within a group,
  observing a pulse is 1.2 times more likely than observing a null.
  \label{tab:groups}}%
\end{table}

\newpage

\subsection*{Profiles around nulls}
\begin{figure}[b]
  \includegraphics[]{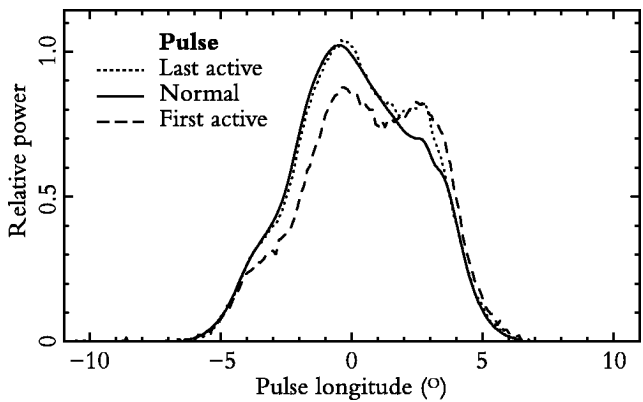}%
  \caption{Average profile around nulls. Black line: average profile
  for normal pulses. Dotted line: average profile for all pulses
  directly before a null. Dashed line: average profile for all pulses
  directly after a null. We have used all (500+) nulls, including
  those in groups.}  \label{img:profiles}
\end{figure}

The next step in investigating how the pulsar turns on and off is to
investigate changes in pulse type or shape before and after
nulls. First, we compare the number of subpulses in normal pulses with
the number observed in pulses that lie in between nulls (cf.\ pulses
25, 26, 143 and 145 in fig. \ref{img:stack0818}). We find that on
average the former contain 1.79$\pm$0.01 subpulses, the latter
1.80$\pm$0.05. To compare pulse shapes, we have averaged the last and
first active pulses around all nulls in our sample, as shown in fig.\
\ref{img:profiles}.

The profile of the first pulse after the null is roughly as expected;
as the pulsar sometimes turns back on only halfway through the pulse
window, we expect the leading edge of the profile to diminish compared
to the trailing edge, and this we indeed see. For the last pulse
before the null one would expect the opposite (the leading edge should
become relatively more important as the pulsar sometimes already turns
off near the end of the null), but this we do not see. The last pulse
before the null actually shows a stronger trailing part, contrary to
what we would expect.

Finally we note that the first and last active pulse more clearly show
two-peaked profiles than the average profile does. As the separation
between the peaks is roughly equal to the subpulse separation P$_2$,
they could be explained by the pulsar turning on and off at certain
preferred subpulse-carousel positions. In this case the subpulses
would not smear out in the average, as in the normal profile, but add
up. We have not been able to verify this suggestion in an independent
way however, so we put it forward with some reservation.

\subsection*{Profiles around low-energy pulses}

\begin{figure}[b]
  \includegraphics[]{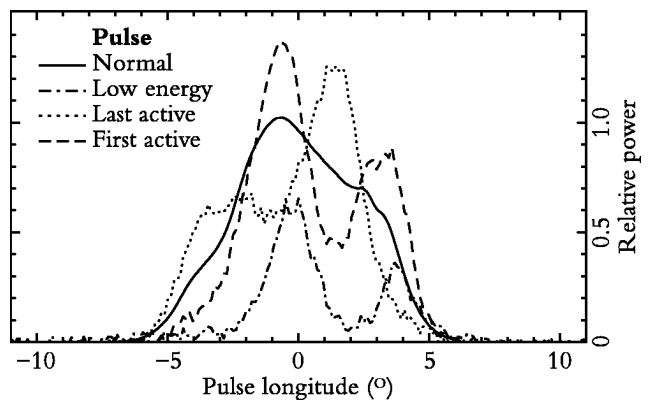}
  \caption{Average profile around lows, as in fig.\
  \ref{img:profiles}. Again, the black line represents the average pulse
  profile. The dotted and dashed lines are the profiles before and
  after the low in this case. The dashed--dotted line shows the
  average profile for the low--energy pulses.}
  \label{img:lowprof}
\end{figure}

As some of the low-energy pulses are probably partial nulls, we also
investigated the profiles of these pulses and the pulses surrounding
them. Similarities between the profiles around nulls and low-energy
pulses, would again suggest they are related.  We averaged the
profiles in and around 80 low-energy pulses, as shown in in fig.\
\ref{img:lowprof}.

First thing to note is the two-peaked profile of the average
low-energy pulses themselves, with the peak separation equal to the
subpulse separation. This indicates that certain subpulse-positions
are overrepresented, a conclusion strengthened by the fact that the
last and first active pulses are offset by the same amount as would be
expected based on normal subpulse drift. Apparently, normal pulses
with certain subpulse positions are less bright than average and
contaminate the equally dim half-null sample.

\begin{figure*}[t]
    \centering
    \includegraphics[width=\textwidth]{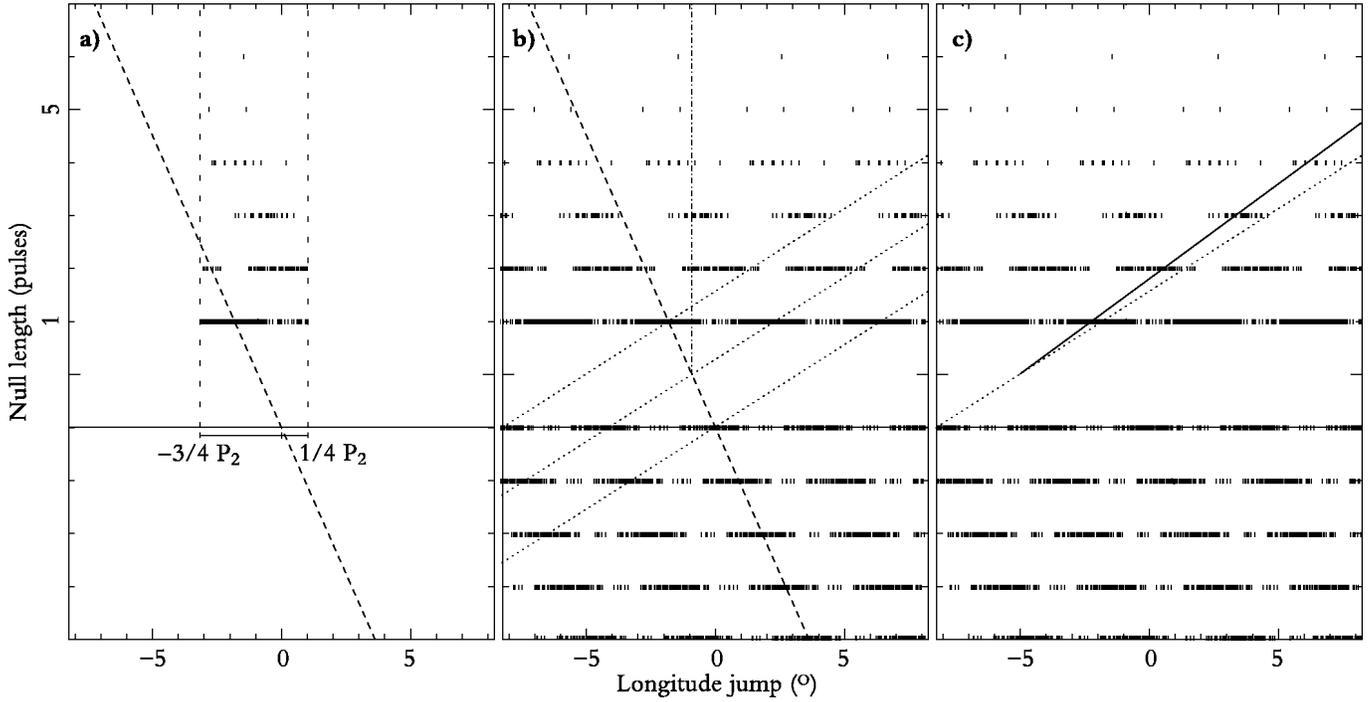}
  \caption{Subpulse-pattern longitude jumps over continuous nulls. Each mark is
  the jump over one null. \newline
  {\bf a)} Mapped into a region from
  $-\frac{3}{4}$P$_2$ to $\frac{1}{4}$P$_2$ around the expected location in
  the continuous-drift case. {\bf b)} The same data, duplicated to show
  subpulse grouping, with simulated pre-null subpulse positions added
  for clarity. The vertical dashed-dotted line predicts subpulse
  positions for a total stop of all drift during the null. The dashed
  lines are the prediction if drifting continuous during the null as it
  did before, for non-aliased drift. The dotted lines show which subpulses
  one spark makes if the drift is aliased. {\bf c)} The same data, now
  with the predicted aliased subpulse path (dotted line), and the
  actual lower driftrate (solid line).}
  \label{img:nulljump}
\end{figure*}

\subsection*{Position jumps over nulls}

Looking at the subpulse drift in PSR~B0818$-$13 in fig.\ 
\ref{img:stack0818}, the drift direction seems obvious (slowly
towards earlier arrival) but in reality it might be different. The
subpulses lack clear, recognizable traits (width, height or spacing)
so instead of moving slowly towards earlier arrival, the subpulses
might actually move more quickly towards later arrival, and still
produce the same image as fig.\  \ref{img:stack0818}. As we only observe
the subpulse position once per pulse period, we undersample the
subpulse movement and cannot determine the underlying speeds
immediately \citep[see also fig.\ 2 in][]{lsr+03}.

By focusing on drift behaviour around nulls, we hope to obtain more
information about the subpulse-drift alias mode and the rotation speed
of the underlying carousel, like we previously did for PSR~B0809+74
\citep{lsr+03}. For each null in our sample, we determine the
positions of the preceding and following subpulses and calculate the
phase jump of the subpulse system over each null.

In fig.\ \ref{img:nulljump}a we plot the longitude jump of the
subpulse system versus null length, for all nulls in our sample. As
the subpulses are indistinguishable, it is not clear which pre-null
subpulse corresponds to which post-null subpulse. Therefore, in fig.\
\ref{img:nulljump}a, all the data is mapped into one region that is
P$_2$ (the average subpulse separation) wide. To visualize the data
better, we have recreated the subpulse pattern by duplicating the
data-points with offsets of \,N$\times$P$_2$ in fig.\
\ref{img:nulljump}b. The real grouping of the subpulses is now much
more apparent. This diagram is analogous to fig.\ \ref{img:stack0818},
but is now thought to show the drifting of the subpulses during the
null, when they are not actually visible. We have also added some
simulated pre-null subpulse positions in the lower half of the
diagram, for clarity.

If, during the null, the subpulse would stop drifting, the subpulses
should follow the vertical dashed-dotted line in fig.\ 
\ref{img:nulljump}b. If the subpulses continue to drift during the
null, they should follow the dashed line. Neither of the two is a
particularly good fit. After the null, the subpulses appear to outrun
the prediction-line, which would indicate the drift-speed has increased
during the null.

\subsection*{The subpulse drift alias mode}

For the dashed line in fig.\ \ref{img:nulljump}b we have assumed that the
drifting is not aliased, i.e.\ that one driftband is made by only one
spark. Another option would be that the carousel is rotating the other
way \citep[as in fig.\ 2b in][]{lsr+03}. In this case, a
single spark does not move towards earlier arrival and follow the
dashed line in fig.\ \ref{img:nulljump}b. It moves towards later arrival
and follows one of the dotted lines.

Compared to these dotted lines of expected positions, the post-null
subpulses are lagging, which means the drifting slows down during a
null. This would be consistent with the behaviour of the only other
well-studied case of drifting-nulling interaction, in
PSR~B0809+74. There the spark-carousel unambiguously decelerates
during a null, even to the point of a total stop. If we assume that
our two-pulsar sample behaves similarly instead of oppositely, the
spark carousels in both should slow down during nulls, and therefore
the real drift direction in PSR~B0818$-$13 must be towards later
arrival. It could do so as illustrated by the dotted lines in fig.\
\ref{img:nulljump}b, but faster too.  If we assume the carousel
consists of 20 sparks, similar to PSR~B0943+10 \citet{dr99} and
PSR~B0809+74 \citep{lsr+03}, we can place an upper limit on the
carousel rotation time. In the slowest aliased case, a
subpulse takes 1.3 pulse periods $P$ to reach the position of its neighbour
(i.e.\ the vertical spacing of the dotted lines in fig.\
\ref{img:nulljump}b). The whole carousel then rotates in {1.3$\,
\times$P$\times$N$_{sparks}\sim$30s}. This already makes
PSR~B0818$-$13 the pulsar with the fastest carousel rotation time
known; if the subpulse drift is more strongly aliased, the
rotation time is even smaller.

\begin{figure*}[t]
   \begin{minipage}[t]{120mm}
   \includegraphics[width=\textwidth]{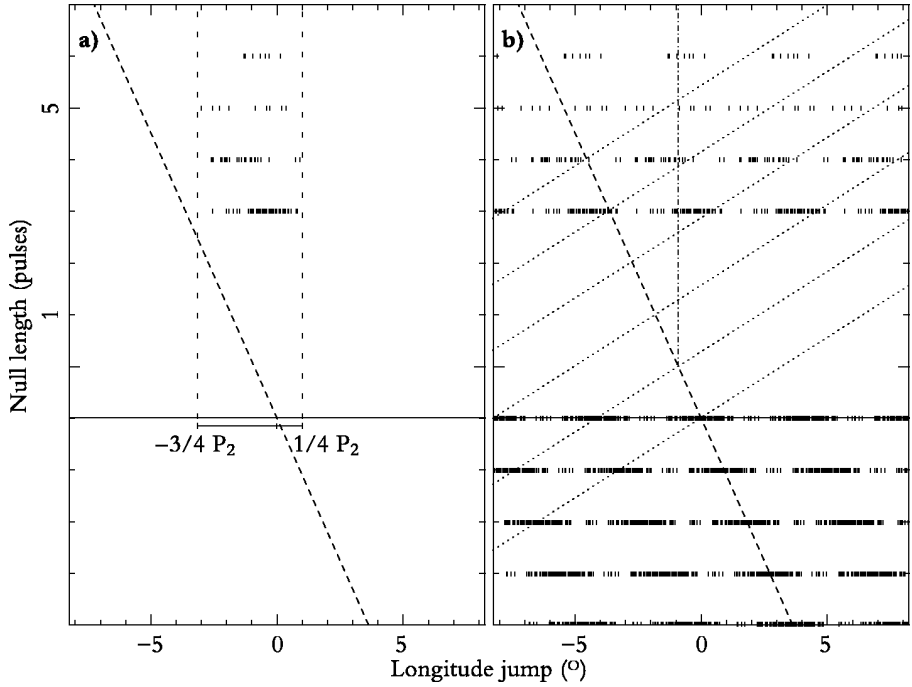}
   \end{minipage}%
   \hspace{5mm}%
   \begin{minipage}[b]{50mm}
   \caption{As fig. \ref{img:nulljump}, but now for intermittent
   nulls.}
   \vspace{1cm}
   \label{img:eventjump}
   \end{minipage}%
\end{figure*}

\subsection*{Subpulse drift during intermittent nulls}

If the carousel in PSR~B0818$-$13 slows down during a null, then why
doesn't it stop like the carousel in PSR~B0809+74 does? Again, the answer
is in the intermittent nature of the apparent long nulls. What appears
to be a long null is actually a series of short nulls: during each of
these the carousel may slow down, only to speed up again in
between. On average, this comes down to a steady lower carousel speed
during nulls. In fig.\ \ref{img:nulljump}, this would translate to a
straight line through the post-null subpulse groups, and this is in
good agreement with the data, as the full line in fig.\ \ref{img:nulljump}c
shows. If we assume the alias mode as illustrated by the dotted line
in  fig.\ \ref{img:nulljump}c, the average driftrate (full line) during a
long apparent null is about 0.87$\pm$0.05 of the normal driftrate
(dotted line). Using the pulse/null ratio of 1.2 we found previously,
this means that during the short nulls that compose this long apparent
null, the drift must be less than half the normal driftspeed to
produce the apparent driftspeed observed.

If this is correct, and the slow drift-rate seen in the apparent long
nulls is due to the subpulse drift starting and stopping every time a
short null changes into a short burst, we must find the same lower
driftrate over the intermittent null-groups. In fig.\
\ref{img:eventjump} we show the subpulse pattern over these
intermittent nulls in the same fashion as fig. \ref{img:eventjump} did
for continuous nulls. Again we note that the subpulses arrive earlier
than expected.  The similarity in figures \ref{img:nulljump} and
\ref{img:eventjump} is striking; even in their drift-rate, the
continuous and intermittent nulls behave identically.

\section{Discussion}
In a rotating-carousel model, the groups of subsequent short nulls and
pulses can be explained in two ways. In the one we have thus far
focussed on, the entire carousel quickly stops and starts emitting. In
this model all subbeams pause at the same time.  In a different model,
only some subbeams in a carousel stop emitting while the carousel
rotation continues, creating pulse-null patterns similar to those
observed. When the off-part of the carousel rotates into view, one
observes a null. One pulse period later, an on-part might have moved
into the pulse window to produce a normal pulse. Such a configuration
would have to meet some very specific criteria, though. If the pattern
is caused by on and off-parts of the carousel turning in and out of
view, one should also observe half-on half-off pulses, with fewer
subbeams. Yet we have found that the pulses in pulse-null groups
contain exactly as many subpulses as normal pulses do. Only a carousel
with many subbeams (roughly several hundreds) at a very high rotation
speed could begin to explain this lack of observed on- to off-part
transitions, but with a subpulse separation of 4$^\circ$, the inferred
size of the total carousel appears unrealistic.

When we compare pulsars B0809+74 and B0818$-$13, the similarities are
striking. Both shine normally for 98-99\% of the time, to be
occasionally interrupted for several seconds and then resume their
normal emission. Both show similar changes in driftrate during nulls.
It seems clear we see the same mechanism at work in these two
pulsars. Yet for PSR~B0809+74 we know the long nulls must be real, as
the subpulse drift stops during these nulls \citep{lsr+03}, while in
PSR~B0818$-$13 we never see long nulls. There we observe periods of
intermittent emission and quietude, in groups about as long as the
nulls in PSR~B0809+74. 

Apparently, the mechanism that causes the emission in PSR~B0809+74 to
fully turn off, only reaches a border-line state in
PSR~B0818$-$13. Why the chances for null or pulse occurrence within
such a state should be almost equal is unclear. The long timescales
involved (even in the intermittent nulls) remains the biggest
problem. In a polar-gap model (the basic physics of which are already
debatable) for example, the potential gap over the magnetic pole can
build up or discharge in microseconds \citep{rs75}. Why the pulsar
would blink on a timescale of seconds, for 10 seconds in total, as
it does in PSR~B0818$-$13 remains unclear; we have not been able to
explain this behaviour with any current pulsar theory. What one can
do, is study the behaviour of different pulsars to find underlying
patterns, which is what we have now done for PSR~B0809+74 and
PSR~B0818$-$13.

The carousel rotation times found in PSR~B0943+10 (41 seconds) and
PSR~B0809+74 ($>$ 200 seconds) pose a similar time-scale problem; in a
simple polar-gap theory, the rotation times are roughly several
seconds \citep{rs75} and the rotation times found thus-far are much
larger. The upper limit of 30 seconds we find for PSR~B0818$-$13
extends the sample, back towards the previously expected values. A
theory like the abovementioned also predicts that the rotation times
scale with pulse period $P$ and magnetic field strength $B$ as
$\frac{B}{P^2}$, which implies that the rotation time of
PSR~B0818$-$13 should be 3 times longer than that of PSR~B0809+74,
instead of the more than 10 times shorter we find. Our determination
of the carousel rotation time in PSR~B0818$-$13 is therefore a step to
finding the dependencies of carousel rotation times on other pulsar
parameters empirically.

\section{Conclusions}
Both the apparently long nulls and the groups of intermittent nulls
can be quantitatively explained by trains of short nulls and pulses.
The subpulse-drift behaviour over the two is identical: during the
nulls, the subpulse drift appears to speed up. Assuming the underlying
sparks slow down, as they do in PSR~B0809+74, we conclude that the
subpulse drift in PSR~B0818$-$13 is aliased. This then leads to a
carousel-rotation time of less than 30 seconds, the fastest one found
to date.

\end{document}